\documentclass[%
 reprint,
superscriptaddress,
 amsmath,amssymb,
 aps,
 pra,
longbibliography
]{revtex4-1}

\newcommand{\be}{\begin{equation}}
\newcommand{\ee}{\end{equation}}
\newcommand{\bea}{\begin{eqnarray}}
\newcommand{\eea}{\end{eqnarray}}
\usepackage{graphicx}
\usepackage{float}
\usepackage{amsmath}
\usepackage{color}
\usepackage{dcolumn}
\usepackage{bm}
\usepackage{hyperref}
\usepackage{braket}
\usepackage{xcolor}
\usepackage{enumerate}
\usepackage{comment}
\usepackage{wasysym}

\begin{document}

\title{Charge and statistics of lattice quasiholes from density measurements:\\ a Tree Tensor Network study}

\author{E. Macaluso}
\affiliation{INO-CNR BEC Center and Dipartimento di Fisica, Universit$\grave{a}$ di Trento, I-38123 Trento, Italy}

\author{T. Comparin}
\affiliation{INO-CNR BEC Center and Dipartimento di Fisica, Universit$\grave{a}$ di Trento, I-38123 Trento, Italy}
\affiliation{Univ Lyon, ENS de Lyon, Univ Claude Bernard, CNRS, 
Laboratoire de Physique, F-69342 Lyon, France}

\author{R. O. Umucal\i lar}
\affiliation{Department of Physics, Mimar Sinan Fine Arts University, 34380 Sisli, Istanbul, Turkey}

\author{M. Gerster}
\affiliation{Institute for Complex Quantum Systems and Center for Integrated Quantum Science and Technologies, Universit$\ddot{a}$t Ulm, D-89069 Ulm, Germany}

\author{S. Montangero}
\affiliation{Dipartimento di Fisica e Astronomia ``G. Galilei'', Universit$\grave{a}$ degli Studi di Padova \& INFN, I-35131 Padova, Italy}

\author{M. Rizzi}
\affiliation{Institute of Quantum Control (PGI-8), Forschungszentrum J$\ddot{u}$lich, D-52425 J$\ddot{u}$lich, Germany}
\affiliation{Institute for Theoretical Physics, University of Cologne, D-50937 K$\ddot{o}$ln, Germany}

\author{I. Carusotto}
\affiliation{INO-CNR BEC Center and Dipartimento di Fisica, Universit$\grave{a}$ di Trento, I-38123 Trento, Italy}

\date{\today}
\begin{abstract}
We numerically investigate the properties of the quasihole excitations above the bosonic fractional Chern insulator state at filling $\nu = 1/2$, in the specific case of the Harper-Hofstadter Hamiltonian with hard-core interactions.
For this purpose we employ a Tree Tensor Network technique, which allows us to study systems with up to $N=18$ particles on a $16 \times 16$ lattice and experiencing an additional harmonic confinement.
First, we observe the quantization of the quasihole charge at fractional values and its robustness against the shape and strength of the impurity potentials used to create and localize such excitations.
Then, we numerically characterize quasihole anyonic statistics by applying a discretized version of the relation connecting the statistics of quasiholes in the lowest Landau level to the depletions they create in the density profile [E. Macaluso \emph{et al.}, \href{https://link.aps.org/doi/10.1103/PhysRevLett.123.266801}{Phys.~Rev.~Lett.~\textbf{123}, 266801}].
Our results give a direct proof of the anyonic statistics for quasiholes of fractional Chern insulators, starting from a realistic Hamiltonian. Moreover, they provide strong indications that this property can be experimentally probed through local density measurements, making our scheme readily applicable in state-of-the-art experiments with ultracold atoms and superconducting qubits.

\end{abstract}

\maketitle

\section{Introduction}

In three spatial dimensions, quantum particles are typically classified into
bosons and fermions, according to the symmetry properties of their many-body
wave functions. In particular, bosonic (fermionic) many-body wave functions must
be globally symmetric (anti-symmetric) in the particle coordinates, meaning that they take an overall $+1$ ($-1$) factor upon particle exchange. This classification
is enriched in two dimensions (2D), where exotic particles called anyons have been predicted to exist~\cite{Wilczek_PRL.49.957, Halperin_PRL.52.1583, Arovas_Wilczek_PRL.53.722, Leinaas_Myrheim_Nuovo_Cimento, Wu_PRL.52.2103, Stern_AoP} so that the effect of particle exchange (resp. braiding) on the many-body wave functions is a generic phase factor $e^{i \varphi_{\text{st}}}$ (resp. $e^{i \varphi_{\text{br}}} = e^{i 2 \varphi_{\text{st}}}$), where the
statistical phase $\varphi_{\text{st}}$ can take any value in $[0,2\pi)$.
While bosons and fermions are characterized by $\varphi_{\text{st}} = 0$ and $\varphi_{\text{st}}=\pi$, Abelian anyons have statistical phase $\varphi_{\text{st}} = \alpha \pi$, with $\alpha$ a non-integer number. In the presence of 
topologically degenerate ground states, the statistical phase factor is further generalized to non-commuting unitary transformations acting on the ground-state manifold, and anyons are said to be non-Abelian~\cite{Moore_NPB.360.2.362, ReadGreen_PRB.61.10267, Ivanov_PRL.86.268, Read_Rezayi_PRB.59.8084, Nayak_RMP.80.1083}.

Among the physical systems for which the existence of anyons has been predicted,
fractional quantum Hall (FQH) systems are probably the most popular ones~\cite{Tsui_Gossard_PRL.48.1559, Laughlin_PRL.50.1395, Moore_NPB.360.2.362, Read_Rezayi_PRB.59.8084, Tong_notes}.
Such strongly correlated quantum fluids can host bulk elementary excitations --called quasiholes (QHs) and quasiparticles (QPs)-- which have been theorized to carry fractional charge and exhibit anyonic behavior.
Although the QH/QP fractional \emph{charge} was measured in electron experiments~\cite{dePicciotto_Nat.389.162}, the anyonic \emph{statistics} of these excitations still lack a clear-cut experimental evidence.
For this reason, a large ongoing effort is based on controllable analog systems, where magnetic quantum-mechanical effects occur for neutrally charged particles such as atoms and photons prepared in the FQH regime~\cite{Cooper_etal_TopoBands_RMP, Ozawa_etal_TopoPhoto_RMP}.

Lattice counterparts of the FQH effect have also attracted strong attention in the recent past.
On the one hand, they include direct generalizations of the FQH effect in 2D lattices~\cite{Sorensen_Lukin_PRL.94.086803, Palmer_Jaksch_PRL.96.180407, Hafezi_Lukin_PRA.76.023613, Bhat_Holland_PRA.76.043601, Kapit_Mueller_PRL.105.215303, Cho_Bose_PRL.101.246809}, generally related to the interacting Harper-Hofstadter (HH) model~\cite{Harper_Proc.Phys.Soc.A.68, Hofstadter_PRB.14.2239}.
On the other hand, inspired by the Haldane model where topologically non-trivial bands appear in the absence of a net external magnetic field~\cite{Haldane_PRL.61.2015}, several other variants have been proposed~\cite{Tang_Wen_PRL.106.236802, Sun_DasSarma_PRL.106.236803, Neupert_Mudry_PRL.106.236804, Sheng_Sheng_NatCom2.389, Regnault_Bernevig_PRX.1.021014, Wu_PRB.85.075116, Grushin_Neupert_PRL.112.156801}.
All these lattice analogs of the FQH states are commonly known as fractional Chern insulators (FCIs)~\cite{Parameswaran_CRP.14.9.816}.

In this context, several theoretical studies investigated the adiabatic preparation of different FCI states and the associated phase diagrams~\cite{Kapit_Simon_PRX.4.031039, Grusdt_Fleischhauer_PRL.113.155301, Barkeshli_Laumann_PRL.115.026802, He_Vishwanath_PRB.96.201103, Motruk_Pollmann_PRB.96.165107, Hudomal_Vasic_PRA.100.053624}. Some works focused on the numerical characterization of these states by inspecting key quantities such as the many-body Chern number, the particle entanglement spectrum, the behavior of the correlation functions and the topological entanglement entropy~\cite{Sterdyniak_Moller_PRB.86.165314, Gerster_Montangero_PRB.96.195123, Rosson_Jacksch_PRA.99.033603}, while others proposed experimentally applicable schemes to identify these elusive strongly correlated phases of matter~\cite{Dong_Pollmann_PRL.121.086401, Repellin_Goldman_PRL.122.166801}. Finally, growing attention has been given to FCI bulk excitations~\cite{Liu_Regnault_PRB.91.045126, Nielsen_Rodriguez_NJP.20.033029, Raciunas_Eckardt_PRA.98.063621, Umucalilar_PRA.98.063629, Jaworowski_Liu_PRB.99.045136, Manna_Nielsen_arXiv.1909.02046}, which (similarly to those characterizing the FQH effect) display fractional charge and anyonic statistics.

In the presence of interactions, finding the exact ground state of a quantum many body problem is often possible only for small system sizes, by means of exact diagonalization (ED) of the Hamiltonian.
A more recent class of numerical techniques is based on the optimization of quantum states in the family of Tensor Network (TN) states, which is a generalization of Matrix Product States -- see for instance Refs.~\cite{Orus_AnnPhys.349.117, Gerster_Montangero_PRB.90.125154, Montangero_intro_TN, Silvi_Montangero_SciPost.10.21468} for general reviews.
Tensor Network states, which are distinguished by their different network topologies, notably include Projected Entangled-Pair States (PEPS)~\cite{Verstraete_Cirac_PRL.96.220601, Schuch_Cirac_PRL.98.140506}, Entangled-Plaquette States (EPS)~\cite{Mezzacapo_NJP_11.083026}, the Multi-scale Entanglement Renormalization Ansatz (MERA)~\cite{Vidal_PRL.99.220405, Vidal_PRL.101.110501}, and the Tree Tensor Network (TTN) states used in this work.
States in the Tensor Network family have already proved their usefulness for treating FCIs in several works, see e.g. in Refs.~\cite{Motruk_Pollmann_PRB.96.165107, Gerster_Montangero_PRB.96.195123, Dong_Pollmann_PRL.121.086401, Rosson_Jacksch_PRA.99.033603}.

In this work we make use of the TTN ansatz used in Ref.~\cite{Gerster_Montangero_PRB.96.195123}, to study the properties (the charge and the statistics) of the QH excitations appearing in the HH model with hardcore bosons.
The key observable of our analysis are the depletions created by the QHs in the density profile of the system. While previous works inspected the QH density depletions mainly to quantify the charge of the QHs and/or to give an estimate of their size [see e.g. in Refs.~\cite{Liu_Regnault_PRB.91.045126, Jaworowski_Liu_PRB.99.045136}], only recently it has been discovered that, for FQH states in the lowest Landau level (LLL), these depletions encode information also on the QH anyonic statistics~\cite{EM_etal_nonAbelian}.
By considering the interacting HH model as a concrete example, here we provide numerical evidence that the experimental protocol proposed in Ref.~\cite{EM_etal_nonAbelian} to extract the QH anyonic statistics from density measurements can be generalized also to FCIs in lattice geometries. Moreover, in order to provide a description which is as close as possible to experimentally realistic situations, we mainly focused our attention on systems with open boundary conditions (OBC) and mesoscopic numbers of particles.
This makes our proposal to measure the anyonic statistics of FCI QHs readily applicable in state-of-the-art experiments with ultracold atoms and superconducting qubits, in which the HH model has already been implemented~\cite{Aidelsburger_NatPhys.11.162, Tai_Greiner_Nat.546.519, Roushan_Martinis_Nat.Phys.3930, Owens_Schuster_PRA.97.013818}, and the occupation of the different lattice sites can be easily measured~\cite{Bakr_Greiner_Nat.462.74, McDonald_Chin_PRX.9.021001, Subhankar_Porto_PRX.9.021002, Roushan_Martinis_Nat.Phys.3930, Ma_Schuster_Nat.566.51}.

The structure of the article is the following: In Sec.~\ref{section:models} we introduce the HH Hamiltonian for bosons with on-site interactions, and we briefly review its basic properties together with the main features of the TTN technique. Then, in the same section, we describe the Monte Carlo sampling of discretized Laughlin wave functions, which we use as an auxiliary method to interpret the results obtained through the TTN ansatz. In Sec.~\ref{section:statistics_from_QH_depletions} we review the main findings of Refs.~\cite{Umucalilar_PRL.120.230403, EM_etal_nonAbelian} and, in particular, the relation connecting the QH braiding phase to the depletions that these 
excitations create in the density profile of continuum systems. This relation is then generalized to the case of lattice systems in Sec.~\ref{section:phase:lattice}.
The main results of our work are reported in Sec.~\ref{section:results}. First, in Sec.~\ref{section:results:charge}, we discuss how to use localized pinning potentials to selectively stabilize states presenting either a single QH or two QHs on top of each other. Then, in Sec.~\ref{section:results:phase}, we provide numerical evidence that the method discussed in Sec.~\ref{section:statistics_from_QH_depletions} is indeed able to reconstruct the QH anyonic statistics even in lattice systems with a relatively small number of particles. Conclusions are finally drawn in Sec.~\ref{section:conclusions}.

\section{Models and Methods}
\label{section:models}

\subsection{Interacting Harper-Hofstadter model and Tree Tensor Networks}
\label{section:models:HH}

The Harper-Hofstadter (HH) model describes non-interacting spinless particles hopping on a two-dimensional square lattice and subjected to an external magnetic field~\cite{Harper_Proc.Phys.Soc.A.68, Hofstadter_PRB.14.2239}.
In the presence of on-site interactions and local energy offsets, the interacting Hamiltonian reads
\begin{equation}
    H = -t \sum_{\langle i, j \rangle} e^{i \theta_{ij}} \hat{a}^{\dagger}_{i} \hat{a}_{j} +\dfrac{U}{2} \sum_{i} \hat{n}_{i} (\hat{n}_{i} -1) +\sum_{i} V_{i} \, \hat{n}_{i} ,
\label{eq:HH_Hamiltonian}
\end{equation}
where $\hat{a}_{i}$ ($\hat{a}^{\dagger}_{i}$) is the bosonic annihilation (creation) operator at the site $i$ of an $L \times L$ square lattice, and where $\hat{n}_{i} = \hat{a}^{\dagger}_{i} \hat{a}_{i}$ is the on-site particle density.
The hopping between neighboring sites $i$ and $j$ is characterized by an amplitude $t$ and by the Peierls phase $\theta_{i j}$, which is related to the magnetic flux passing through the system~\cite{Peierls_ZP.80.11.1993}.
\begin{figure}[b]
\centering
\includegraphics[width=0.45\textwidth]{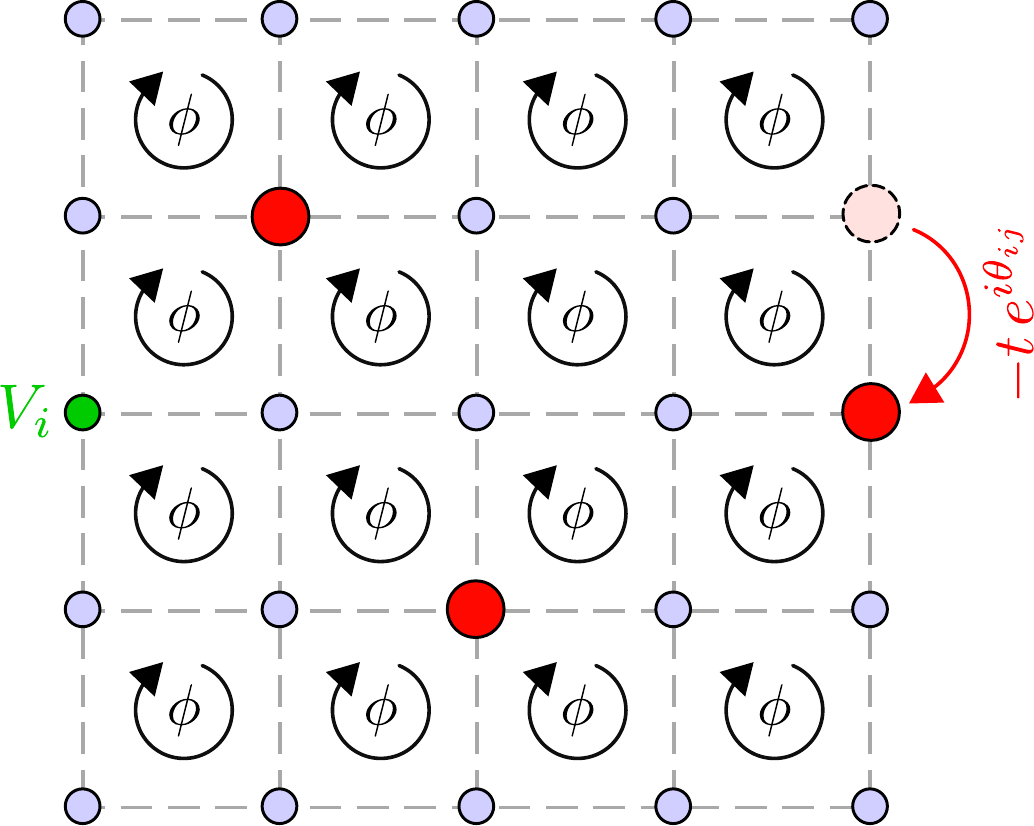}
\caption{
Schematic representation of three bosons (red circles) on a $5\times 5$ lattice with $\phi = \alpha \Phi_{0}$ flux per plaquette.
}
\label{fig:Harper_Hofstadter_model}
\end{figure}
We focus on the case of a uniform magnetic flux per plaquette $\phi = \alpha \Phi_{0}$, where $\Phi_{0} = 2 \pi \hbar /e$ is the magnetic flux quantum.
For a given value of the flux density $\alpha$, multiple choices of the phases $\theta_{ij}$ are possible. Among them, we choose the Landau gauge considered in Ref.~\cite{Gerster_Montangero_PRB.96.195123}.
Moreover, we consider the limit of hard-core bosons (that is, with infinite on-site repulsion $U$), where each site can host at most one particle. The local energy offsets $V_{i}$ represent additional attractive or repulsive potentials on some lattice sites [see Fig.~\ref{fig:Harper_Hofstadter_model}], which can encode the localized impurity potentials to pin the QHs and/or an additional trapping potential for the particles. Even though in the following we focus only on harmonic confinements, more complicated forms could be studied with no additional difficulties.

Our study of this model is based on the Tree Tensor Network technique.
This is a variational ansatz in the class of Tensor Network states, with some noticeable advantages and shortcomings compared to other TN ansatz states, all essentially related to their loop-free structure. On the one hand, TTNs do not capture the entanglement area law for arbitrarily large systems (which, instead, PEPS would do by construction). On the other hand, their simple connectivity allows for very efficient computational algorithms~\cite{Shi_Vidal_PRA.74.022320, Montangero_intro_TN, Silvi_Montangero_SciPost.10.21468}. Thus, the bond dimension $D$ can be pushed to large enough values to compensate for the intrinsic weaknesses of the ansatz, yielding reliable numerical results for system sizes which are beyond reach with exact diagonalization.

As shown in Ref.~\cite{Gerster_Montangero_PRB.96.195123}, the TTN technique is capable to fully reproduce the properties of a FCI state without QH excitations, in the interacting Harper-Hofstadter model. These properties include the correct topological degeneracy when the system is placed on a torus, the many-body Chern number, the behavior of correlation functions, and the entanglement-entropy scaling laws. For the case of a FCI state with QH excitations, we benchmark the validity of the TTN ansatz through a comparison to exact results -- see Appendix~\ref{app:ED}.

The core results of this work are obtained for systems on a $16 \times 16$ square lattice, with open boundary conditions (OBC) and in the presence of an additional harmonic trapping.
In the absence of trapping potentials, the number of magnetic fluxes would have to be commensurate with the number of particles in order to obtain an appropriate magnetic filling $\nu = N / N_{\Phi}$, where the total number of fluxes for systems with OBCs is $N_{\Phi} = (L-1)^2 \alpha$.
However, adding a harmonic confinement relaxes this constraint, and for a given number of particles $N$ it is possible to obtain a FCI state as the ground state for different values of the flux density $\alpha$~\footnote{Clearly, if the target state is a FCI state at magnetic filling $\nu$, also in the presence of a harmonic confinement the ratio between the number of particles and the total number of flux quanta piercing the system cannot exceed the value of $\nu$.}. Note that different values of $\alpha$ correspond to different ratios between the lattice constant $a$ and the effective (model dependent) magnetic length $l_{B} \equiv \sqrt{A/ 2 \pi}$, where $A$ is the area of the magnetic unit cell~\cite{Liu_Regnault_PRB.91.045126}. In particular, for the model under study (i.e., the HH model on a square lattice), $A = a^2/\alpha$ and such a ratio reads
\begin{equation}
\dfrac{a}{l_\textrm{B}} = \sqrt{2 \pi \alpha} \, .
\label{eq:HH:abylB}
\end{equation}
This expression of the lattice spacing in terms of the effective magnetic length will be of crucial importance both in Sec.~\ref{section:models:discrete_laughlin} and for the analysis of the QH density depletions in Sec.~\ref{section:results:phase}.

A final remark concerns the competition between the different possible states (FCIs, superfluid states, Mott insulators, charge-density-waves, etc.) which might be the ground state of the HH Hamiltonian in Eq.~\eqref{eq:HH_Hamiltonian} -- see for instance Refs.~\cite{Umucalilar_Mueller_PRA.81.053628, Natu_DasSarma_PRA.93.063610, Hugel_Pollet_PRB.96.054431, Bai_Angom_PRA.98.023606}. In this respect, superimposing an additional harmonic confinement to the lattice further enriches an already complicated scenario.
However, for all the values of $\alpha$ we consider in the following it is possible to identify a finite interval of strengths of the harmonic confinement in which the system ground state is a FCI state.
Although a detailed analysis of the different competing phases as a function of the confining strength is an interesting task, we postpone it to a future work and to focus here only on intermediate values of the harmonic potential strength for which the system ground state has a FCI nature.

\subsection{Monte Carlo sampling of discretized Laughlin-like wave functions}
\label{section:models:discrete_laughlin}

As we will see, the TTN approach allows us to capture a very close approximation of the true ground state of the Harper-Hofstadter model for sizes compatible with near-future experiments in cold gases or circuit QED systems. However, there is a practical limit on the lattice size and particle number that can be treated.

To address more technical questions related to the effect of the lattice grid, rather than of the system size [see Sec.~\ref{section:phase:lattice}], we employ an auxiliary method which provides more flexibility.
This consists in the Monte Carlo sampling of a discretized version of the Laughlin-like wave functions~\cite{Laughlin_PRL.50.1395}, which are evaluated on the sites of a two-dimensional grid [as done for instance in Ref.~\cite{Hafezi_Lukin_PRA.76.023613}].
More explicitly, at magnetic filling $1/2$, the wave function for $N$ particles and $k$ QHs reads
\begin{equation}
    \Psi_{k\text{QH}}(z_1, \dots, z_N) =
    \left[\prod^{N}_{i=1} \prod^{k}_{\mu=1} \left(z_{i} - \eta_{\mu} \right) \right] \Psi_{L}(z_1, \dots, z_N)
\label{eq:Laughlin_QH_wf}
\end{equation}
where $\eta_{\mu}$ is the position of the $\mu$-th QH, $\Psi_{L}(z_1, \dots, z_N)$ is the celebrated Laughlin wave function, i.e.
\begin{equation}
    \Psi_{L}(z_1, \dots, z_N) = \prod^{N}_{i<j} (z_i - z_j)^2 \, e^{- \sum^{N}_{i} |z_{i}|^{2} / 4 l^{2}_{\text{B}}} ,
\label{eq:Laughlin_wf}
\end{equation}
and, on a lattice, the position of the $j$-th particle ($z_j = x_j + i y_j$, with the complex coordinate notation) only takes discrete values (that is, with $x_j/a$ and $y_j/a$ integer).

The use of the Laughlin wave function as a reference state for Harper-Hofstadter systems in the limit of low flux density $\alpha$ is common in literature~\cite{Sorensen_Lukin_PRL.94.086803, Hafezi_Lukin_PRA.76.023613}.
In this work, we use the wave functions in Eq.~\eqref{eq:Laughlin_QH_wf} to study the properties of QH excitations, and to provide a comparison with the HH model.
The usefulness of this comparison relies on the study in Ref.~\cite{Liu_Regnault_PRB.91.045126}, which proves that the density profile around FCI QHs is a discretized version of the one of Laughlin QHs in continuum space, up to a proper rescaling of the length units.
Of course, for a given flux density $\alpha$, a proper comparison with the HH model requires that the lattice spacing of the discretization grid satisfies Eq.~\eqref{eq:HH:abylB}.

For a discretized Laughlin-like wave function, it is straightforward to generalize the Monte Carlo technique which is typically used to extract observable quantities for the continuum-space Laughlin state~\cite{Morf_Halperin_PRB.33.2221, Comparin_Laughlin_Metropolis}. This consists in sampling configurations $\lbrace z_1, \dots, z_N \rbrace$ distributed as in $|\Psi(z_1, \dots, z_N)|^2$, which give access to observables like the density profile on the lattice. By changing the discretization of $\Psi$ (that is, by tuning the lattice constant $a$), we have access to different values of $a/l_\mathrm{B}$. This is valuable to have a one-to-one comparison with HH systems [where this ratio is set by $\alpha$ -- see Eq.~\eqref{eq:HH:abylB}], but also to perform a more systematic study of the convergence towards the continuum limit ($a/l_\mathrm{B} \to 0$).

Note, to conclude, that other choices are available for the discretized version of Laughlin-like wave functions, see for instance Ref.~\cite{Nielsen_PRB.91.041106}, and we expect that similar results can be obtained for all of them.

\section{Anyonic statistics from density-profile measurements}
\label{section:statistics_from_QH_depletions}

We now consider FQH and FCI states of $N$ particles, in the presence of some localized QH excitations. We start by reviewing a recent proposal to characterize the braiding phase of these excitations through density profile measurements in the continuum~\cite{Umucalilar_PRL.120.230403, EM_etal_nonAbelian}, and we then conjecture on its generalization to the lattice case.

\subsection{Continuum systems}
\label{section:phase:continuum}

In Refs.~\cite{Umucalilar_PRL.120.230403, EM_etal_nonAbelian}, some of the current authors developed a scheme to access the anyonic statistics for QH excitations of FQH states. In contrast to other proposals based on interferometric experiments [see for instance Ref.~\cite{ParedesZoller_PRL.87.010402}], this proposal only requires static measurements. More precisely, a useful relation was obtained to relate the QH braiding phase $\varphi_\text{br}$ and the expectation value $\langle \hat{L}_z \rangle$ of the angular momentum operator taken on some specific quantum states.
Each state hosts two QHs at positions $\vec{\eta}_1$ and $\vec{\eta}_2$, which are either diametrically opposite with respect to the system center or on top of each other.
For FQH states in the LLL, the correspondence between $\langle \hat{L}_z \rangle$ and the mean square radius $\langle r^2 \rangle$~\cite{Ho_Mueller_PRL.89.050401} simplifies our scheme even further, since it implies that one can obtain the braiding phase simply by measuring the density profile in the presence of QHs.
More precisely, $\langle \hat{L}_z \rangle /\hbar + N = N \langle r^{2} \rangle / 2 l^{2}_{\text{B}}$ and the QH braiding phase reads~\cite{Umucalilar_PRL.120.230403}
\begin{equation}
	\dfrac{\varphi_{\text{br}}}{2\pi} = \dfrac{N}{2 l^{2}_{\text{B}}} \left[ \langle r^{2} \rangle_{\text{opp}} - \langle r^{2} \rangle_{\text{over}} \right ] .
	\label{eq:phi_br_1}
\end{equation}
The subscripts of $\langle r^2 \rangle$ refer to two aforementioned QH configurations that need to be considered: ``opp'' indicates diametrically opposite QHs (i.e., $\vec{\eta}_{1} = - \vec{\eta}_{2}$), while ``over'' indicates overlapping QHs (i.e., $\vec{\eta}_{1} = \vec{\eta}_{2}$).
We stress that, in Eq.~\eqref{eq:phi_br_1}, $l_{\text{B}}$ is the actual magnetic length $l_{\text{B}} \equiv \sqrt{\hbar/q B}$, depending on the particle charge $q$ and on the magnetic field $B$, and not a model dependent quantity.

Several practical issues appear when applying Eq.~\eqref{eq:phi_br_1}, due to the fact that it involves global properties of the FQH cloud:
First, one has to measure $\langle r^{2} \rangle_{\text{opp}}$ and $\langle r^{2} \rangle_{\text{over}}$ on systems with exactly the same number of particles $N$ and with the same value of $|\vec{\eta}_1|$ and $|\vec{\eta}_2|$.
Second, one needs to consider large-enough system sizes, so that $\vec{\eta}_1$ and $\vec{\eta}_2$ (in the ``opp'' configuration) are far enough from each other and from the cloud boundaries.
Third, global properties of the system, like $\langle r^{2} \rangle$, are not robust with respect to low-energy perturbations, which in the case of pinned QHs typically consist in the excitation of edge modes~\cite{EMIC_PRA.96.043607, EMIC_PRA.98.013605}.

These three issues are drastically mitigated once we rewrite Eq.~\eqref{eq:phi_br_1} in terms of the depletion $d(\vec{\rho})$, which is the change in the FQH density profile at position $\vec{r}$ induced by a QH at position $\vec{\eta}$, with $\vec{r} = \vec{\eta} + \vec{\rho}$.
More precisely, we define the depletion profiles $d_{1\text{QH}}(\vec{\rho})$ and $d_{2\text{QH}}(\vec{\rho})$ close to single or double QH as
\begin{equation}
d_{k\text{QH}}(\vec{\rho}) =
n_{0\text{QH}}(\vec{r}) -
n_{k\text{QH}}(\vec{r}),
\qquad k=1,2.
\label{eq:depl_cont}
\end{equation}
Here $n_{k\text{QH}}(\vec{r})$ represents the average density on a state with $k$ QHs at position $\vec{\eta}$, for $k\in\lbrace 0, 1, 2 \rbrace$.
By computing $d_{1\text{QH}}(\vec{\rho})$ and $d_{2\text{QH}}(\vec{\rho})$, the braiding phase can be expressed as~\cite{EM_etal_nonAbelian}:
\begin{equation}
	\dfrac{\varphi_{\text{br}}}{2\pi} = \dfrac{1}{2 l^{2}_{\text{B}}} \int d\vec{\rho}  \, \rho^{2} \left[ d_{2\text{QH}} (\vec{\rho}) - 2 d_{1\text{QH}} (\vec{\rho}) \right].
	\label{eq:phi_br_2}
\end{equation}
For finite-size systems the integral is defined up to a cutoff distance $|\vec{\rho}| = R_{\text{max}}$, which should be significantly larger than the QH size, but also small enough to avoid spurious effects due to the FQH cloud boundaries.
This guarantees the mathematical equivalence between Eqs.~\eqref{eq:phi_br_1} and \eqref{eq:phi_br_2}. 
The dependence of Eq.~\eqref{eq:phi_br_2} on the cutoff $R_{\text{max}}$ is characterized by damped oscillations which converge towards the actual value of $\varphi_\text{br}$~\cite{EM_etal_nonAbelian}, as visible in Figs.~\ref{fig:depl_and_phi_CaseI}(b) and \ref{fig:depl_and_phi_CaseII} (c) and (d).

The new expression for the braiding phase, Eq.~\eqref{eq:phi_br_2}, has some clear advantages over Eq.~\eqref{eq:phi_br_1}. First, it only depends on the local density perturbation induced by the QHs, rather than on the global shape of the cloud. This simplifies the measurement, which is now independent on the precise position of the QHs.
Second, the constraint on the cutoff distance $R_{\text{max}}$ is milder than the one on the distance $|\vec{\eta}_1 - \vec{\eta}_2|$ in the ``opp'' configuration, since now there is no need to consider two spatially separated QHs.
Third, local properties like the depletions in Eq.~\eqref{eq:phi_br_2} are not modified by perturbations which are localized at the edge of the system, so that this measurement is expected to be more robust against edge excitations and finite-temperature effects.

In previous works, we numerically confirmed the validity of Eqs.~\eqref{eq:phi_br_1} and \eqref{eq:phi_br_2} by a Monte Carlo study of two paradigmatic FQH states. We considered the Laughlin~\cite{Umucalilar_PRL.120.230403} and Moore-Read~\cite{EM_etal_nonAbelian} wave functions, and for the latter we focused on the case of two non-Abelian QHs~\cite{Moore_NPB.360.2.362, Nayak_RMP.80.1083}. Nevertheless, the procedure is totally general, and it could be applied to any state in the LLL.

\subsection{Lattice generalization}
\label{section:phase:lattice}
While in the previous works we presented the theory relating the anyonic statistics of QHs to their depletion profiles in a continuum geometry, here we discuss how these relations [Eqs.~\eqref{eq:phi_br_1} and \eqref{eq:phi_br_2}] change if one considers lattice systems.

The mean square radius $\langle r^{2} \rangle$ now reads
\begin{equation}
\langle r^{2} \rangle =
\dfrac{1}{N} \sum_{j}  \, \langle \hat{n}_{j} \rangle  \, |\vec{r}_j|^2
	,
\label{eq:msr_lattice}
\end{equation}
where $\vec{r}_j$ is the position of the $j$-th site.
With this definition, it is straightforward to generalize Eq.~\eqref{eq:phi_br_1} to the lattice case:
\begin{equation}
   \dfrac{\varphi_{\text{br}}}{2\pi} = \dfrac{1}{2 l^{2}_{\text{B}}} \sum_{j} \left[ \langle \hat{n}_{j} \rangle_{\text{opp}} - \langle \hat{n}_{j} \rangle_{\text{over}} \right ] |\vec{r}_j|^2  .
   \label{eq:phi_msr_lattice}
\end{equation}
Similarly, we define the depletions $d_{1\text{QH}}(\vec{\rho}_j)$ and $d_{2\text{QH}}(\vec{\rho}_j)$ as
\begin{equation}
d_{k\text{QH}}(\vec{\rho}_j) =
\langle \hat{n}_j \rangle_{0\text{QH}} -
\langle \hat{n}_j \rangle_{k\text{QH}},
\qquad k=1,2,
\label{eq:depl_lattice}
\end{equation}
where $\langle \hat{n}_j \rangle_{k\text{QH}}$ is the average density on site $j$ for a state with $k$ QHs at position $\vec{\eta}$, and where $\vec{r}_j = \vec{\eta} + \vec{\rho}_j$.
Thus Eq.~\eqref{eq:phi_br_2} becomes
\begin{equation}
\dfrac{\varphi_{\text{br}}}{2 \pi} =
\dfrac{1}{2 l^{2}_{\text{B}}}
\sum_{j}
\left[ d_{2\text{QH}}(\vec{\rho}_j) - 2 d_{1\text{QH}}(\vec{\rho}_j) \right] |\vec{\rho}_j|^2 ,
\label{eq:phi_depl_lattice}
\end{equation}
where the sum over $j$ is restricted to sites with $|\vec{\rho}_j| < R_\mathrm{max}$, as for Eq.~\eqref{eq:phi_br_2}.
Note that both expressions for the braiding phase $\varphi_{\text{br}}$ [see Eqs.~\eqref{eq:phi_msr_lattice} and \eqref{eq:phi_depl_lattice}] explicitly depend on the \emph{effective} magnetic length $l_{\text{B}}$, defined in Eq.~\eqref{eq:HH:abylB}.

Before moving on, we need to stress that the angular momentum operator is not properly defined on a lattice, and therefore the relation between the QH braiding phase and the density profile [Eqs.~\eqref{eq:phi_msr_lattice} and \eqref{eq:phi_depl_lattice}] is not mathematically rigorous in this case. However, the idea of generalizing Eqs.~\eqref{eq:phi_br_1} and \eqref{eq:phi_br_2} to the case of FCIs on a lattice is motivated (and partially justified) by two observations:
First, the study by Liu and co-authors~\cite{Liu_Regnault_PRB.91.045126} clearly shows that, for systems with periodic boundary conditions (PBCs), the density profile close to a single FCI QH is a discrete sampling of the continuum case, once a suitable (model-dependent) effective magnetic length is introduced. In Sec.~\ref{section:results}, we explicitly confirm this result to the case of the HH model with OBC, both for a single and a double QH.
Second, we perform a complementary analysis of the lattice case, based on the discretized Laughlin wave function described in Section~\ref{section:models:discrete_laughlin}. This flexible ansatz allows us to scan different values of the grid spacing $a$, and to compute the braiding phase through the discretized versions of Eqs.~\eqref{eq:phi_br_1} and \eqref{eq:phi_br_2}. In Table~\ref{table:discrete_laughlin:phi_br}, we report the numerical results obtained via Eq.~\eqref{eq:phi_msr_lattice}, while we will use Eq.~\eqref{eq:phi_depl_lattice} in Section~\ref{section:results:phase}. We find that the braiding phase of the QH excitations of the discretized Laughlin state is in full agreement --up to some small deviations due to finite-size and discretization effects-- with the expected value $\varphi_\mathrm{br}/(2\pi) = 1/2$, within the statistical uncertainties of the Monte Carlo method. It is remarkable that this still holds true for $a / l_\mathrm{B} \simeq 1.77$, which
corresponds to the maximum flux density that can be realized in the HH model, i.e. $\alpha = 1/2$.

\begin{table}[h]
\begin{center}
\begin{tabular}{ c c c}
$a/l_\mathrm{B}$ & $\varphi_\mathrm{br}/(2\pi)$ & $\alpha$ \\
\hline
0.5605 & $0.53 \pm 0.04$  & 0.05 \\
0.9708 & $0.55 \pm 0.04$  & 0.15 \\
1.2533 & $0.57 \pm 0.04$  & 0.25 \\
1.7725 & $0.56 \pm 0.04$  & 0.50  \\
\end{tabular}
\end{center}
\caption{
Quasihole braiding phase $\varphi_\mathrm{br}$ for a discretized Laughlin-like wave function (with $N=40$ particles), for different values of the grid spacing $a$. Numerical results (listed in the second column, with their statistical uncertainty) are obtained via Eq.~\eqref{eq:phi_msr_lattice} and Monte Carlo sampling of the discretized Laughlin wave functions.
In the third column, we list the value of $\alpha$ that would correspond, in the HH model, to the chosen lattice spacing $a/l_\mathrm{B}$ [see Eq.~\eqref{eq:HH:abylB}].}
\label{table:discrete_laughlin:phi_br}
\end{table}

In the next section we verify the validity of Eq.~\eqref{eq:phi_depl_lattice} on the ground states of the interacting HH model, obtained with the TTN technique.

\section{Results}
\label{section:results}

In our study of the HH model through the TTN ansatz, we consider two sets of parameters: $N=12$ bosons and $\alpha=0.15$ (Case I), or $N=18$ particles and $\alpha=0.25$ (Case II). These two choices for $\alpha$ correspond to $a/l_\mathrm{B} \simeq 0.97$ and $a / l_\mathrm{B} \simeq 1.25$, respectively [see Eq.~\eqref{eq:HH:abylB}]. 
Considering different values of $\alpha$ allows us to modify $N$ in a significant way without changing the size of the lattice, and it also gives us the opportunity to inspect discretization and flux-dependent effects.
Note also that for Case II we choose $\alpha=0.25$, which is one of the most appealing flux densities for realizing almost flat bands in realistic experiments~\cite{Aidelsburger_NatPhys.11.162, Tai_Greiner_Nat.546.519, Roushan_Martinis_Nat.Phys.3930, Owens_Schuster_PRA.97.013818}.
We also introduce an additional harmonic confinement in the form $V_j = \Omega |\vec{r}_j - \vec{r}_\mathrm{center}|^2$, where the trap center corresponds to the center of the $L\times L$ lattice.
The value of $\Omega$ cannot be chosen arbitrarily, since it is not guaranteed that the ground state is an FCI for \emph{any} trapping strength. The main FCI signature is the formation of a flat-density region in the center of the trap (with density equal to $\alpha/2$), which signals its incompressibility.
Through the TTN technique, we identify a set of $\Omega$'s where the density profile shows an incompressible central region with density $\alpha/2$. In Case I, this occurs at least for $\Omega/t$ between $10^{-4}$ and $3 \times 10^{-3}$, while for Case II this happens for $\Omega/t$ between $2 \times 10^{-3}$ and $10^{-2}$. In the following, we choose trapping strengths within these intervals, namely $\Omega/t = 1 \times 10^{-3}$ for Case I and $\Omega/t = 3\times10^{-3}$ for Case II.
A systematic study of the stability of the incompressible core as a function of both the magnetic flux density $\alpha$ and the confining strength $\Omega$ is certainly interesting for experimental purposes, but it is left for future work.

Concerning the TTN ansatz, we use bond dimensions as large as $D = 500$ (for the largest $N$), and we verify that the systematic error in the relevant observables is negligible [see Appendix~\ref{app:TTN_convergence}].

The two sets of parameters used in the numerical calculations are summarized in Table~\ref{table:HH_parameters}.
In the following, we describe how to stabilize QH states for these two cases and we apply the procedure described in Section~\ref{section:phase:lattice} to demonstrate their anyonic nature.
\begin{table}[h]
\centering
\begin{tabular}{c|c c c c}
& $N$ & $L$ & $\alpha$ & $\Omega/t$ \\
 \hline
Case I & 12 & 16 & $0.15$ & $1 \times 10^{-3}$ \\
Case II & 18 & 16 & $0.25$ & $3 \times 10^{-3}$ \\
\end{tabular}
\caption{Parameters of the Harper-Hofstadter model used in this work.}
\label{table:HH_parameters}
\end{table}

\subsection{Stabilization and charge of the quasiholes}
\label{section:results:charge}
As a preliminary step before entering into the discussion of the braiding phase, we show that it is possible to stabilize states with either one or two (overlapping) QHs. To do so, we make use of some localized pinning potentials with different strengths and shapes (e.g. point-like potentials acting on a single-site, plus-shaped potentials, square-shaped potentials, etc.).
For systems with PBCs, it is already known that some of these potentials can select states with a single QH as the ground state of the system~\cite{Liu_Regnault_PRB.91.045126, Umucalilar_PRA.98.063629, Jaworowski_Liu_PRB.99.045136}. On the other hand, only recently Ra$\check{\text{c}}$i$\bar{\text{u}}$nas and co-workers have inspected the stabilization of QHs in systems with OBCs~\cite{Raciunas_Eckardt_PRA.98.063621}.
However, the use of exact diagonalization techniques limited their analysis to systems with small number of particles ($N=4$), preventing them from clearly observing the expected QH fractional charge.

\begin{figure}[tb]
\centering
\includegraphics[width=0.98\linewidth]{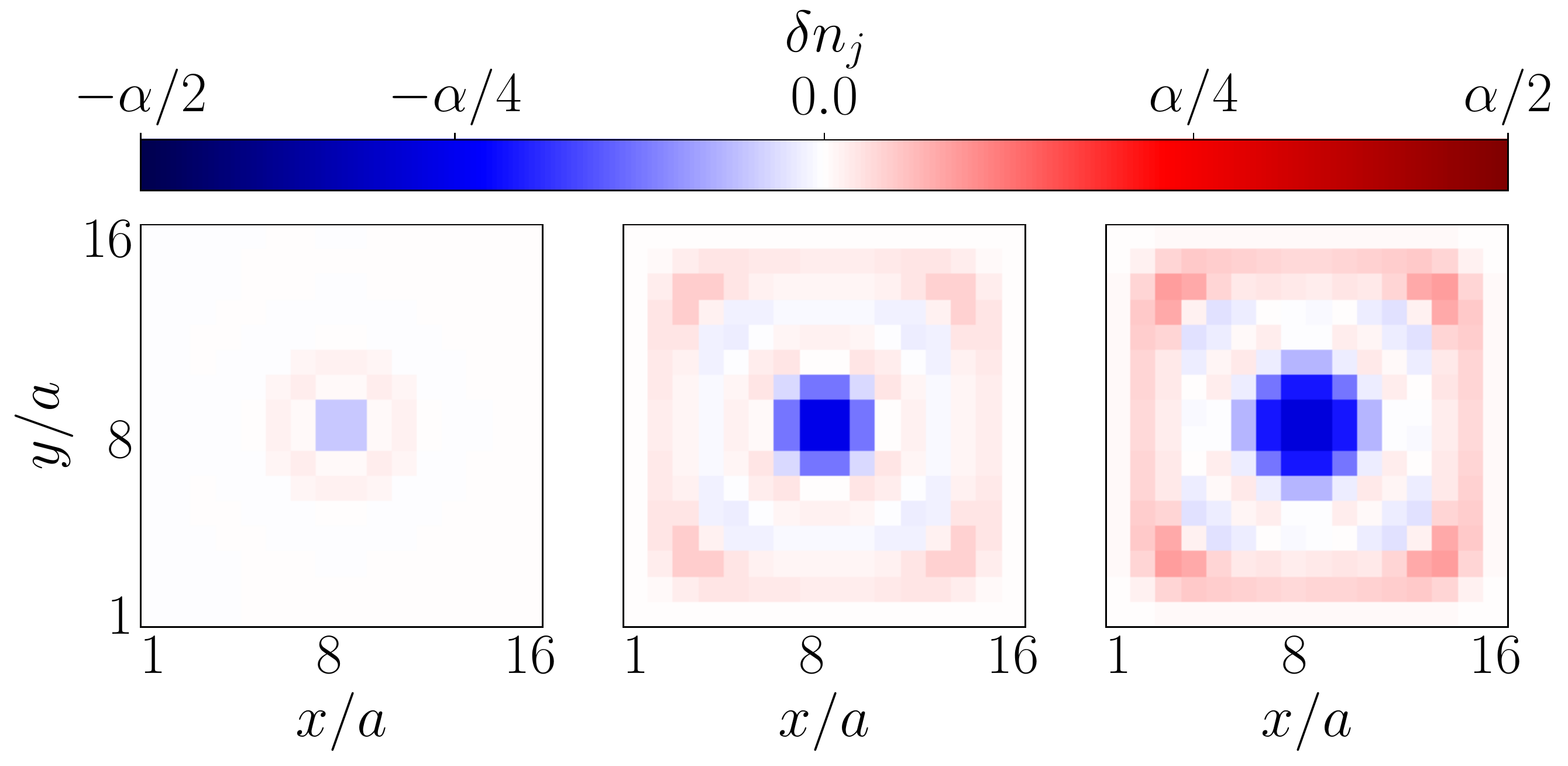}
\vskip 0.15cm
\includegraphics[width=0.98\linewidth]{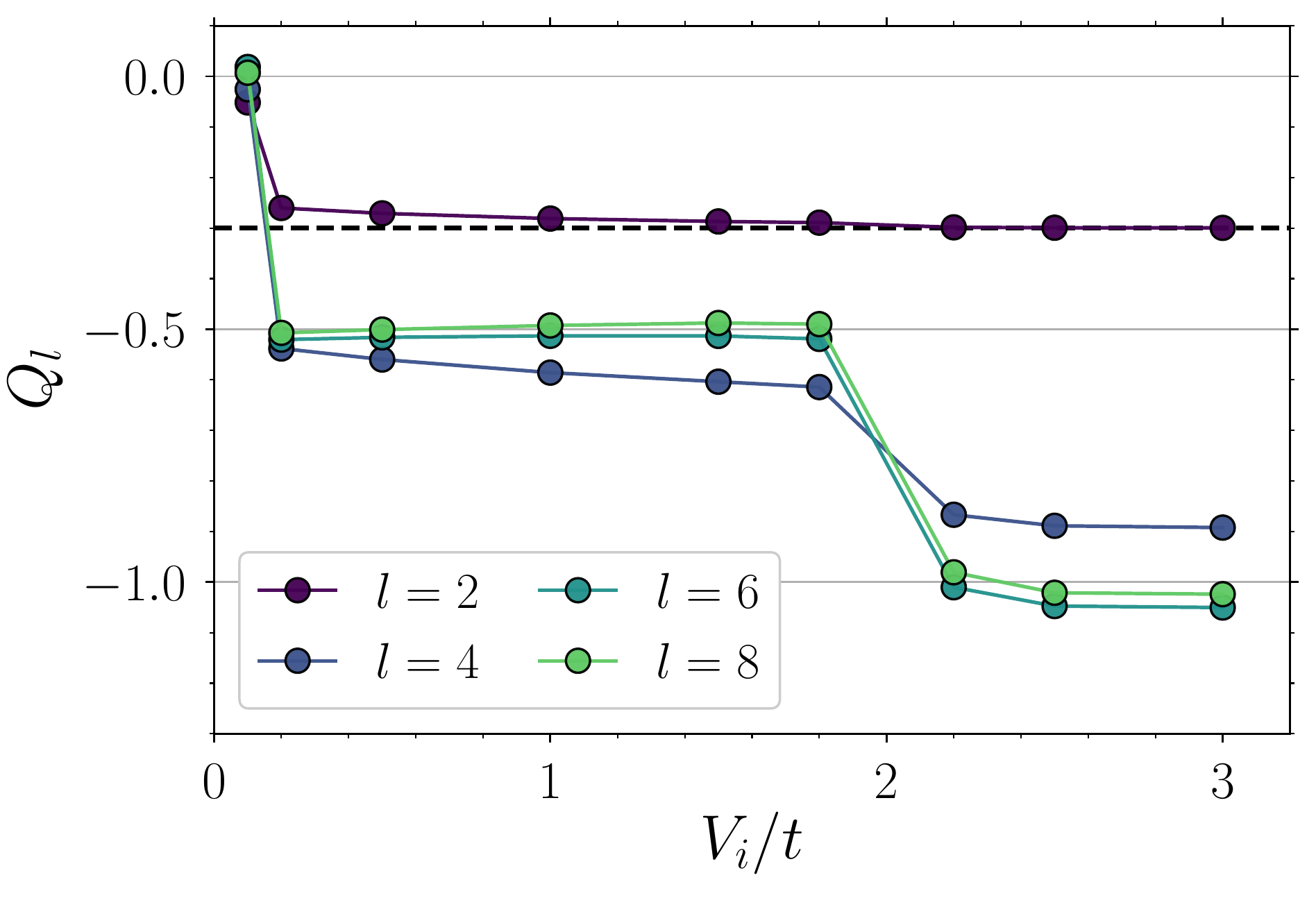}
\caption{Upper panel: variation in the occupation of the different lattice sites with respect to the unperturbed state $\delta n_{j} = \langle \hat{n}_j \rangle_{V_i} - \langle \hat{n}_j \rangle_{V_i =0}$, induced by a $2 \times 2$ pinning potential of strength $V_{i}$ ($V_{i}/t = 0.1, \, 1.0$, and $2.5$, from left to right) placed in the center of the lattice. The system parameters considered here are those of Case I -- see Table~\ref{table:HH_parameters}.
Lower panel: Charge $Q_l$ induced by a square-shaped pinning potential with intensity $V_i$ on the central plaquette (i.e. on the central $2 \times 2$ lattice sites). For large enough $l$, the charge saturates to the expected values for one and two QHs, namely $Q_l=-1/2$ and $Q_l=-1$. 
The other parameters are as in Case I -- see Table~\ref{table:HH_parameters}. The charge is computed as in Eq.~\eqref{eq:QH_charge}, for the values of $l$ indicated in the legend. The horizontal dashed line corresponds to $-4 \times (\alpha/2) = -0.3$, that is, minus four times the bulk density of the unperturbed FCI state. The TTN bond dimension is $D=350$ for all data points.}
\label{fig:charge}
\end{figure}

\begin{figure*}
\centering
\includegraphics[width=0.98\textwidth]{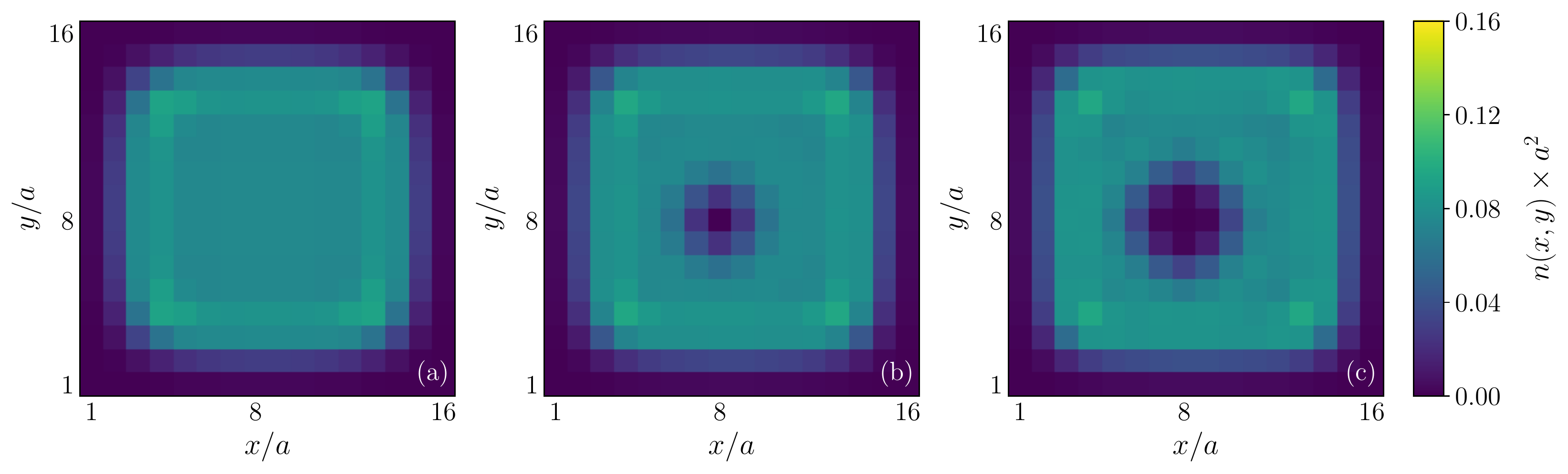}
\includegraphics[width=0.98\textwidth]{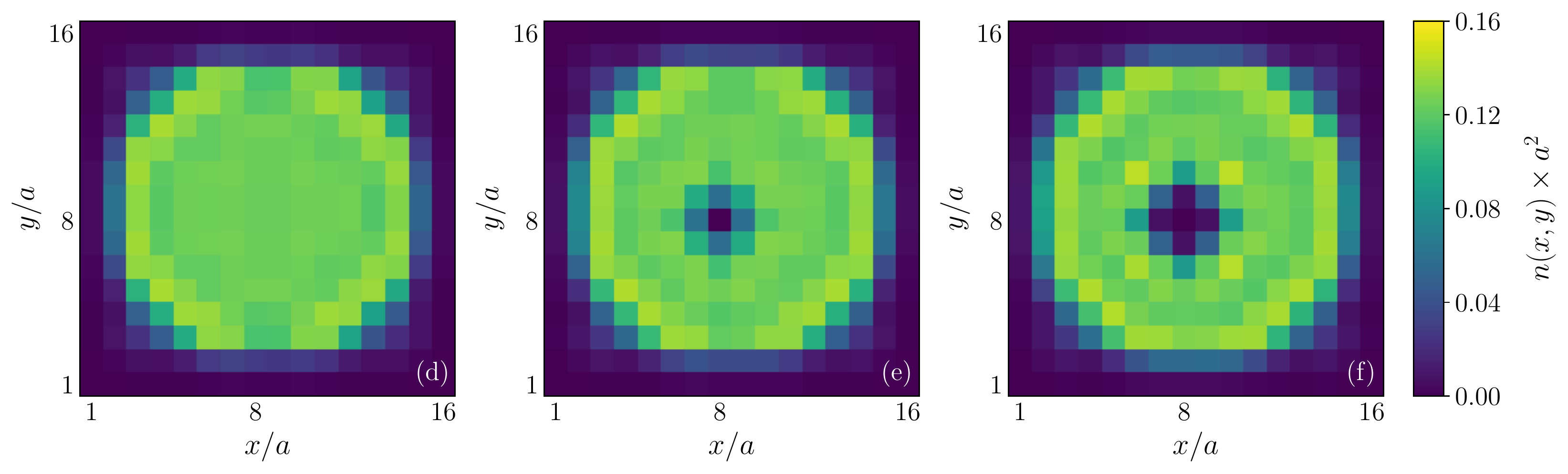}
\caption{Two-dimensional density profiles for Case I [panels (a), (b) and (c)] and Case II [panels (d), (e) and (f)], obtained by TTN calculations. Panels from left to right represent states with zero, one or two QHs pinned at site $(8,8)$. For Case I we used a single-site pinning potential of strength $V_{i}/t=2.0$ located at $(8,8)$, to pin the single QH [panel (b)], and a plus-shaped potential of strength $V_{i}/t=1.0$ acting on the sites $(7,8)$, $(8,7)$, $(8,8)$, $(8,8)$ and $(9,8)$, to pin the two overlapping QHs [panel (c)]. For Case II we used a single-site pinning potential of strength $V_{i}/t=4.0$, still at $(8,8)$, for the single QH [panel (e)], and the same plus-shaped potential as before, for the two overlapping QHs [panel (f)]. The TTN bond dimension is $D=350$ for Case I and $D=500$ for Case II.}
\label{fig:2D_densities_TTN}
\end{figure*}

We consider the Hamiltonian in Eq.~\eqref{eq:HH_Hamiltonian} with Case I parameters and we add a pinning potential with finite intensity $V_i$ on each of the four sites of a plaquette at the center of the trap (squared-shaped potential). Then we compute the charge $Q_l$, defined as:
\begin{equation}
Q_l = \sum_{j} \delta n_{j} = \sum_{j} \,\big[ \langle \hat{n}_j \rangle_{V_i} - \langle \hat{n}_j \rangle_{V_i =0} \big],
\label{eq:QH_charge}
\end{equation}
where the sum is restricted to the sites $j$ of a $ l \times l$ square at the center of the trap. For large $V_i$, the density on the central $2 \times 2$ plaquette tends to zero, so that (for $l=2$) $Q_l$ tends to $-4\times(\alpha/2) = -0.3$, which is minus four times the bulk density in the absence of pinning potentials. 
This is clearly visible in the lower panel of Fig.~\ref{fig:charge}.
Note that for $\alpha = 0.25$ the plateau in $Q_2$ would occur exactly at $-1/2$, independently on the number of QHs pinned by the impurity potential, and this could lead to the misinterpretation of the data.
The correct values for the (fractional) charge of the QH excitations are recovered for large enough $l$.
The lower panel in Fig.~\ref{fig:charge} clearly shows that, by increasing the potential strength, $Q_l$ saturates at $-1/2$ and $-1$, which are the expected charges for one and two QHs, respectively.
Note that the exact location of the transitions between 0, 1 and 2 QHs also depends on the trapping strength $\Omega$, on the flux density $\alpha$, and on the particle number $N$.

Along this line, there are two important remarks we want to make. 
The first one concerns the robustness of the bulk against weak perturbations: While for pinning potentials strong enough to create some QH excitations, the density depleted by the QHs is displaced to the system boundary [see middle and right pictures in Fig.~\ref{fig:charge}, top panel], weak repulsive potentials modify the systems density only around the potential position, giving an overall depleted charge $Q_{l} \simeq 0$ for sufficiently large values of $l$ and no accumulated charge on the boundary [see left picture in Fig.~\ref{fig:charge}, top panel].
Note also that the robustness of the bulk is expected to increase with the number of particles, since in the large-$N$ limit the density depleted by the QH must be pushed much farther away to reach the system boundary.

The second point concerns the role played by harmonic confinement, which we found to be very important to efficiently stabilize the QH states.
The advantage of introducing an additional harmonic trap is twofold: On the one hand, it shifts the edge modes of the FCI state to higher energies. On the other hand, it allows the system to automatically regulate its spatial extension to minimize the energy.
Thus the system can adapt itself to the presence of the pinning potentials by putting QHs in correspondence with the potentials and displacing the extra density to the boundary, without the need of removing particles or modifying the flux density $\alpha$.
In other words, in the presence of a harmonic confinement the system automatically chooses the right amount of flux quanta in order to be an FCI at filling $\nu=1/2$ with the most suitable number of QH excitations.
Note that this behavior is completely different from what usually happens in systems with OBCs (without additional trapping potentials), where the total number $N_{\text{QH}}$ of QH excitations is set by the fact that $\nu = (N + N_{\text{QH}}/2)/(L-1)^{2} \alpha$.

Even though we considered squared-shaped potentials in our analysis of the QH charge, we verified that similar results can be obtained for pinning potentials of different forms [see for instance the single-site and the plus-shaped potentials used in Sec.~\ref{section:results:phase} to stabilize one and two QHs, respectively]. The discussion about the pros and cons of the different pinning potentials, and the effects they have on the QH properties, is postponed to the next section.

\subsection{Quasihole statistics}
\label{section:results:phase}
As discussed in Sec.~\ref{section:phase:continuum}, the method proposed in Ref.~\cite{EM_etal_nonAbelian} has the great advantage of allowing us to determine the QH braiding phase in a setup that would be prohibitive for the more traditional schemes involving real braiding and interference. By lifting the needs of an explicit rotation of the QHs, and of their spatial separation, this method allows us to work with trapped samples with OBCs (rather than with abstract periodic structures) also when the number of particles is relatively small~\footnote{The presence of the system boundary dramatically reduces the available flat-density region, where the QHs must be pinned (far from each other) and eventually braided to address their statistical properties through the traditional methods.}.
On this basis, we decided to apply the lattice generalization of that protocol --see Sec.~\ref{section:phase:lattice}-- to the FCI QHs under study [see parameter sets in Table~\ref{table:HH_parameters}]. We stress that both in Case I and in Case II, the system size is too small to accommodate two QHs far enough from each other and from the system boundary to apply the usual braiding schemes.

The key ingredients to measure the QH braiding phase, through the protocol described by Eq.~\eqref{eq:phi_depl_lattice}, are the density depletions $d_{1\text{QH}}(\vec{\rho}_j)$ and $d_{2\text{QH}}(\vec{\rho}_j)$.
To take advantage of Eq.~\eqref{eq:phi_depl_lattice}, the set of vectors $\{ \vec{\rho}_j \}$, i.e. the distances between the lattice sites and the QH center, must be the same for both $d_{1\text{QH}}$ and $d_{2\text{QH}}$. As a consequence, both the single QH and the two overlapping QHs must be centered at the same position. To satisfy this constraint we identified two possible configurations: (i) a single-site potential to pin the single QH and a plus-shaped potential (centered on the same site and spread over five sites) to pin the two overlapping QHs; (ii) a $2 \times 2$ square-shaped potential with different strengths to selectively pin one or two QHs.
\begin{table}[bt]
\centering
\begin{tabular}{c | c | c | c}
Marker & Parameters & $1$QH pinning & $2$QH pinning \\
 \hline
\Circle & Case I & single-site     & plus-shaped      \\
        &        & ($V_i/t = 2.0$) & ($V_i/t = 1.0$)  \\
$\Box$ & Case II     & single-site     & plus-shaped   \\
       &             & ($V_i/t = 4.0$) & ($V_i/t = 0.7$)  \\
$\Diamond$ & Case II    & single-site     & plus-shaped   \\
           &            & ($V_i/t = 4.0$) & ($V_i/t = 1.0$)  \\
$\hexagon$ & Case II    & square-shaped     & square-shaped   \\
           &            & ($V_i/t = 0.4$)   & ($V_i/t = 2.0$)  \\
\end{tabular}
\caption{Hamiltonian and pinning potential parameters considered in the TTN calculations of the quasihole braiding phase [see Figs.~\ref{fig:depl_and_phi_CaseI} and \ref{fig:depl_and_phi_CaseII}].}
\label{table:markers}
\end{table}

\begin{figure}[tb]
\centering
\includegraphics[width=0.48\textwidth]{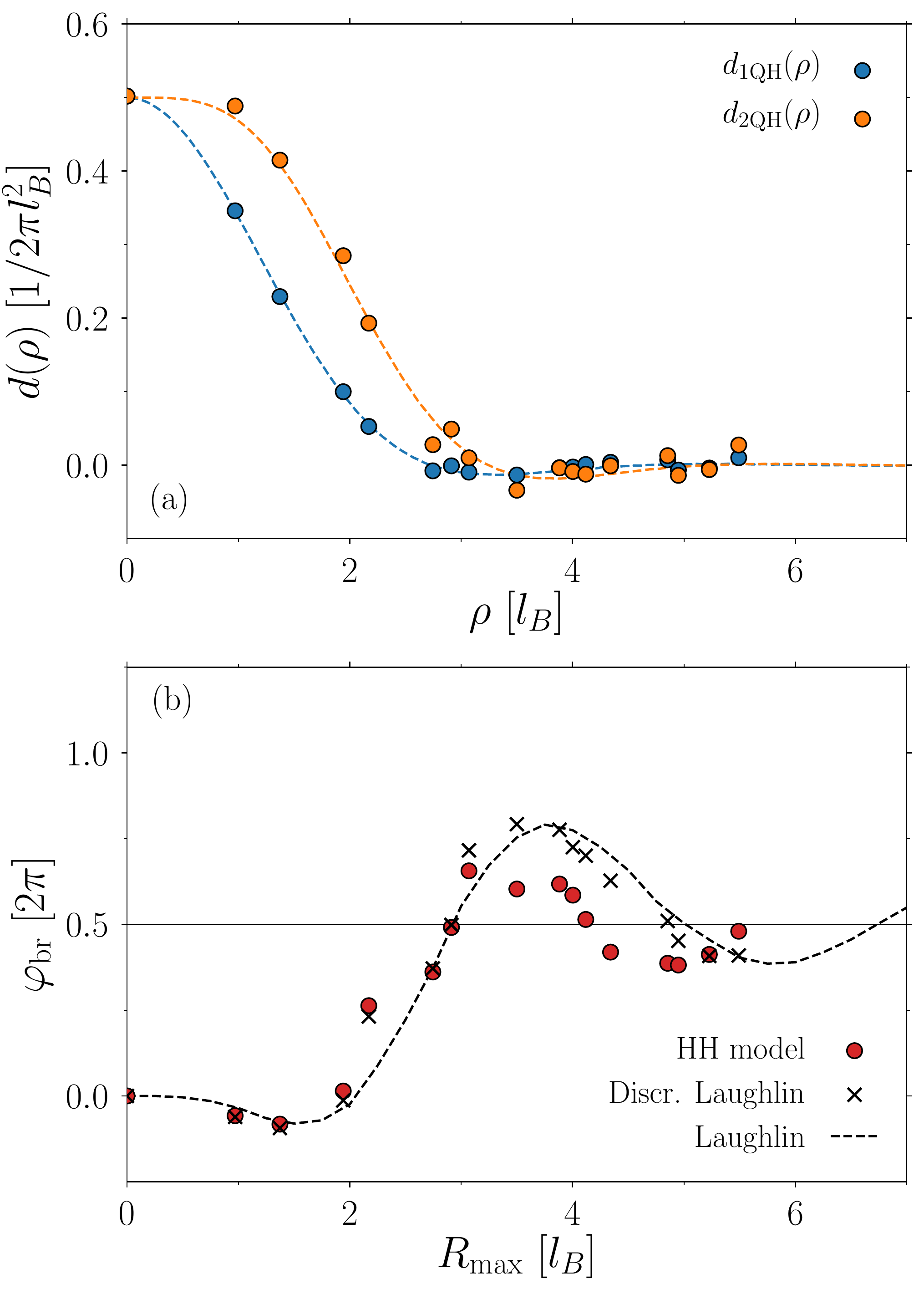}
\caption{Panel (a): Comparison between the density depletions created by a single QH (blue dots) and two overlapping QHs (orange dots) in the interacting HH model and in the continuum $\nu=1/2$ Laughlin state (blue and orange dashed lines). Panel (b): Comparison between the quasihole braiding phase, as a function of the cutoff radius $R_{\text{max}}$, for the interacting HH model (red dots), for the discretized quasihole wave functions in Eq.~\eqref{eq:Laughlin_QH_wf} with $a \simeq 0.97 \, l_{\text{B}}$ (black crosses), and for their continuum version (black dashed line). The Hamiltonian parameters are those of Case I and the properties --shape and strength-- of the pinning potentials considered in the TTN calculations are reported in Table~\ref{table:markers}. The TTN bond dimension is $D=350$ for all data sets.}
\label{fig:depl_and_phi_CaseI}
\end{figure}

\begin{figure*}
\includegraphics[width=0.96\textwidth]{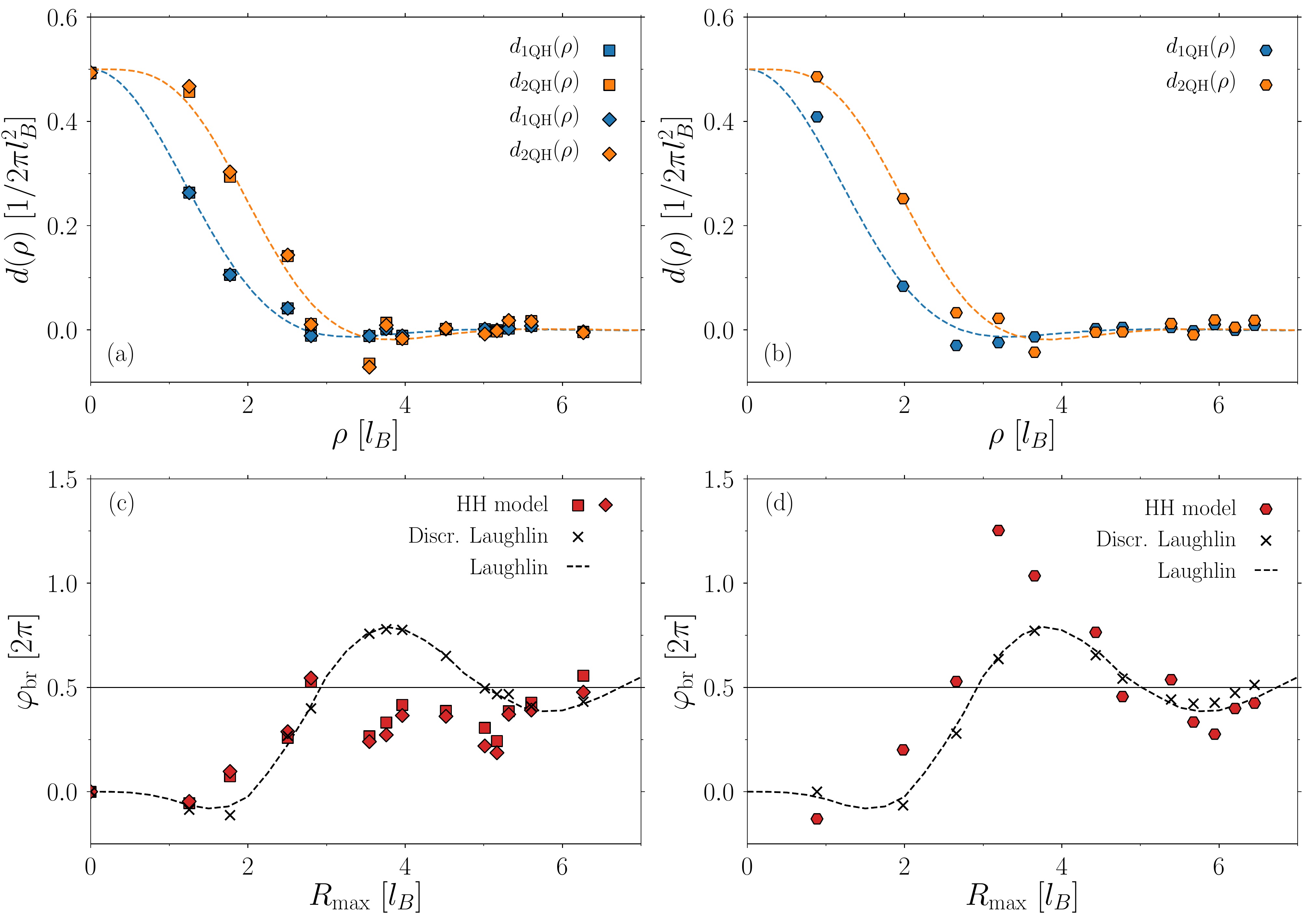}
\caption{Panels (a) and (b): Comparison between the density depletions created by a single QH (blue markers) and two overlapping QHs (orange markers) in the interacting HH model and in the continuum $\nu=1/2$ Laughlin state (blue and orange dashed lines). Different markers indicate the use of different pinning potentials [see Table~\ref{table:markers}] to create the quasiholes in a system with Case II parameters. Panels (c) and (d): Comparison between the quasihole braiding phase, as a function of the cutoff radius $R_{\text{max}}$, obtained for the interacting HH model with different pinning potentials (red markers), for the discretized quasihole wave functions in Eq.~\eqref{eq:Laughlin_QH_wf} with $a \simeq 1.25 \, l_{\text{B}}$ (black crosses), and for their continuum version (black dashed line). In panels (a) and (c) quasiholes are centered on a lattice site, while in (b) and (d) they are located at the center of a plaquette. The TTN bond dimension is $D=500$ for all data sets.}
\label{fig:depl_and_phi_CaseII}
\end{figure*}

In order to extract the depletions $d_{1\text{QH}}(\vec{\rho}_j)$ and $d_{2\text{QH}}(\vec{\rho}_j)$, one needs the density profiles $\langle \hat{n}_{j} \rangle_{k\text{QH}}$ characterizing states without QHs ($k=0$), with a single QH ($k=1$), and with $k=2$ overlapping QHs [see Eq.~\eqref{eq:depl_lattice}]. We evaluated these density profiles by means of the TTN technique [see examples in Fig.~\ref{fig:2D_densities_TTN}], and the resulting QH density depletions $d_{1\text{QH}}(\vec{\rho}_j)$ and $d_{2\text{QH}}(\vec{\rho}_j)$ are shown in panel (a) of Fig.~\ref{fig:depl_and_phi_CaseI} and in panels (a) and (b) of Fig.~\ref{fig:depl_and_phi_CaseII}.

To simplify the interpretation of the figures, we use the following marker notation:
First, blue (orange) markers denote the data obtained through the TTN technique for the depletion due to a single (double) QH.
Second, the different marker shapes are associated with different physical parameters and/or pinning potentials, as summarized in Table~\ref{table:markers}.
Finally, regarding the results for the QH braiding phase [panel (b) of Fig.~\ref{fig:depl_and_phi_CaseI} and panels (c) and (d) of Fig.~\ref{fig:depl_and_phi_CaseII}], red markers refer to the TTN results for the interacting HH model, the black crosses indicate the data obtained through the sampling of the discretized Laughlin-like wave functions, while the black dashed line corresponds to the behavior of $\varphi_{\text{br}}$ for the Laughlin QHs in the continuum.

The comparison of the QH density depletions obtained through the TTN technique with those characterizing the Laughlin QHs is in general very good [see panel (a) in Fig.~\ref{fig:depl_and_phi_CaseI} and panels (a) and (b) in Fig.~\ref{fig:depl_and_phi_CaseII}], up to small deviations.
At large distances there are discrepancies due to finite-size effects which affect both $d_{1\text{QH}}(\vec{\rho}_j)$ and $d_{2\text{QH}}(\vec{\rho}_j)$.
To be precise, these are caused by the presence of the density bump close to system boundary, which modifies the tail of the depletions~\footnote{Note that the density bump close to the system boundary is not a particular feature of the lattice system under study, but it is common to all FQH states in geometries with OBCs.}. 
To mitigate this problem, we restrict the plotted data to distances for which the depletions behave smoothly and the QH charges are close to the expected values (i.e., $-0.5$ and $-1$ for the single and double QH, respectively).
At small and intermediate distances, instead, the reason why the QH density depletions appear slightly different with respect to their continuum counterparts is twofold.
On the one hand, there is the effect of the pinning potentials [see for instance the radial profiles of the 2QH density depletions reported in panel (a) of Fig.~\ref{fig:depl_and_phi_CaseII}], which is more evident for $\alpha=0.25$ than for $\alpha=0.15$.
This dependence on the flux density is due to the fact that larger $\alpha$'s correspond to larger values of the ratio $a/l_{\text{B}}$.
This implies that, in the case of $\alpha=0.25$, the same (finite-width) pinning potential used for $\alpha=0.15$ effectively acts at larger distances from the center of the QH, where the system density is higher and the potential can have stronger effects on the QH depletions $d_{1\text{QH}}$ and $d_{2\text{QH}}$.
On the other hand, there are unexpected deformations which are present also in the density depletion of the single QH pinned by a single-site potential [see blue squares and blue diamonds in panel (a) of Fig.~\ref{fig:depl_and_phi_CaseII} at short and intermediate distances]. Zero-range pinning potentials should not deform the QH density depletions and therefore we consider the observed discrepancies as model-dependent effects, probably due to the specific properties of the HH bands for a given value of $\alpha$.
This would explain why this behavior is more evident for Case II ($\alpha=0.25$) than for Case I ($\alpha=0.15$).

Finally, we used the method described in Sec.~\ref{section:phase:lattice} to extract the braiding phase $\varphi_{\text{br}}$ of the lattice QHs, as a function of the cutoff radius $R_{\text{max}}$ [see Figs.~\ref{fig:depl_and_phi_CaseI} (b) and \ref{fig:depl_and_phi_CaseII} (c) and (d)].
We compare it with the results obtained for both the QHs of the continuum-space Laughlin state and their discretized counterparts.
Note that for the sampling of the discretized QH wave functions we chose a different discretization grid spacing for Case I and Case II, since they are characterized by a different value of the ratio $a/l_{\text{B}}$ [see Eq.~\eqref{eq:HH:abylB}].

For Case I, the behavior of $\varphi_{\text{br}}$ obtained for the QHs of the interacting HH model is extremely similar to the one predicted by the Monte Carlo sampling of the discretized Laughlin QHs up to a certain value of $R_{\text{max}}$ [see Fig.~\ref{fig:depl_and_phi_CaseI}, panel (b)]. After that, at larger cutoff radii, we observe small deviations between the two data sets, reflecting the deformations in the depletion $d_{2\text{QH}}(\vec{\rho}_j)$ caused by the plus-shaped pinning potential.
However, the results shown in Fig.~\ref{fig:depl_and_phi_CaseI}(b) clearly indicate that for Case I the anyonic nature of the QHs in the interacting HH model can be probed through simple density measurements.

The interpretation of the QH braiding phase obtained for Case II is more subtle.
As we have already discussed, for $\alpha = 0.25$ the density depletions of the lattice QHs display more visible discrepancies, with respect to their continuum counterparts, than for $\alpha=0.15$.
Moreover, due to the $|\vec{\rho}_{j}|^2$ factor in Eq.~\eqref{eq:phi_depl_lattice}, these discrepancies in the depletions translate into even stronger effects affecting the behavior of $\varphi_{\text{br}}(R_{\text{max}})$.
This is clearly visible in panel (c) of Fig.~\ref{fig:depl_and_phi_CaseII}.
Despite discrepancies in the profile of the density depletions, the method proposed in Eq.~\eqref{eq:phi_depl_lattice} is still valid and the correct results for $\varphi_{\text{br}}(R_{\text{max}})$ are recovered for large enough integration regions --where the deformations in the density depletions are damped.
We attribute this behavior to the topological robustness~\footnote{The robustness of Laughlin properties on the lattice has been already seen by explicit computation of many-body Chern number~\cite{Sorensen_Lukin_PRL.94.086803, Gerster_Montangero_PRB.96.195123}.} of the QH braiding properties and we expect further confirmation of this interpretation from future studies of of larger systems. In spite of these additional difficulties, the results obtained for Case II confirm that the anyonic nature of the QHs can be inspected through simple density-profile measurements also in the experimentally promising $\alpha=0.25$ case.

Even though our method to measure QH braiding phase seems to be robust against the deformations induced by extended pinning potentials in the depletion profiles, they affect the behavior of $\varphi_{\text{br}}(R_{\text{max}})$ at small and intermediate values of the cutoff radius.
To get rid of these effects, one could in principle create multiple QHs at the same position by locally inserting the suitable amount of flux quanta, in the presence of a single-site potential able to pin a single QH~\cite{Grusdt_Fleischhauer_PRL.113.155301, Wang_Eckardt_PRL.120.243602, Raciunas_Eckardt_PRA.98.063621}.
Despite being experimentally feasible, in the theoretical framework this flux-insertion procedure requires time-dependent simulations of the interacting HH Hamiltonian, which at the moment go beyond the capabilities of our TTN technique.
Along this line, a recent application of the TTN ansatz for time-dependent simulations opens interesting perspectives~\cite{Kohn_Montangero_arxiv.1907.00009}.

\section{Conclusions and Outlook}
\label{section:conclusions}
In this work we used a Tree Tensor Network ansatz to study the properties of the quasihole excitations of the fractional Chern insulator described by the Harper-Hofstadter Hamiltonian with hardcore interactions. The loop-free geometry of the Tree Tensor Network ansatz allowed us to study systems with open boundary conditions, far beyond the typical system sizes manageable by exact diagonalization calculations.

In this way, first we showed that it is possible to use localized pinning potentials to stabilize states hosting either a single or two overlapping quasiholes and that the expected fractional charge of these excitations is already clearly visible in a $N = 12$ particle system.
In this respect, we discovered that superimposing an additional harmonic confinement to the lattice greatly simplifies the stabilization of the quasihole states.

Then, to characterize the statistics of the quasiholes, we applied a lattice version of the equation proposed in Ref.~\cite{EM_etal_nonAbelian} and relating the quasihole braiding phase to the depletions induced by such excitations in the system density.
Our results clearly show that these excitations are anyons, namely that they are neither bosons nor fermions (for which $\varphi_{\text{br}} = 0, 2 \pi$), and that their braiding phase is very close to the predicted one, i.e. $\varphi_{\text{br}} = 2 \pi \nu$.
In spite of the obvious limitations in the accuracy of the measurement of $\varphi_{\text{br}}$, mainly due to the size of the state-of-the-art samples, we stress that our results have been obtained for systems which are too small to accommodate two spatially separated quasiholes and adiabatically braid them to inspect their statistical properties, as it would be required by traditional measurement schemes.

As a result, the present study provides numerical evidence that the anyonic statistics can indeed be observed through simple density measurements in state-of-the-art experiments with ultracold atoms and superconducting qubits. First of all, the flux densities we considered in this work --i.e., $\alpha = 0.15$ and $0.25$-- are already within the current experimental capabilities.
The case of $\alpha=0.25$, in particular, is of great interest from the experimental point of view.
At such flux density the single-particle spectrum of the HH Hamiltonian is characterized by four energy bands, with a very convenient (low) ratio between the width of the lowest band and its separation from the higher ones. This makes $\alpha = 0.25$ one of the most appealing flux densities for realizing almost flat bands in realistic experiments~\cite{Aidelsburger_NatPhys.11.162, Tai_Greiner_Nat.546.519, Roushan_Martinis_Nat.Phys.3930, Owens_Schuster_PRA.97.013818}.
At the same time, the lattice size we looked at --i.e, $L = 16$-- is comparable with the one used in Ref.~\cite{Owens_Schuster_PRA.97.013818}.
On top of that, adding an overall harmonic confinement to the lattice should be possible in both relevant setups: For ultracold atomic systems the harmonic trap is typically already present on top of the lattice potential (and in most cases it is difficult to remove); while platforms based on superconducting qubits should allow one to independently tune the different on-site potentials $V_{i}$~\cite{Chen_Martinis_PRL.113.220502, Roushan_Martinis_Nat.Phys.3930}.
Finally, the hard-core constraint describes well the on-site interactions in superconducting qubits~\cite{Roushan_Martinis_Nat.Phys.3930}. 
This condition might be difficult to reach with ultracold atoms; however, as suggested in Refs.~\cite{Raciunas_Eckardt_PRA.98.063621, Rosson_Jacksch_PRA.99.033603, Hudomal_Vasic_PRA.100.053624}, we believe that considering finite-strength interactions instead of the hard-core ones should not modify our results in a considerable way.
A comprehensive analysis of the case of soft-core interactions is left for a future work. 

From the theoretical point of view, our work, combined with the observations by Liu and co-workers~\cite{Liu_Regnault_PRB.91.045126}, numerically shows that the expression relating the QH braiding phase to the QH density depletions is valid also for more general lattice systems, once a suitable effective magnetic length is introduced.
Since on a lattice the angular momentum operator is not properly
defined, such an expression can not be rigorously derived as done in Ref.~\cite{EM_etal_nonAbelian} for the continuous system.
This opens the interesting question whether it is possible to find a deeper and more general explanation of the link between the statistical properties of the QHs and their density profiles, which seem to be independent of the model under study.

Possible extensions of this work include other FCI states, and in particular those hosting non-Abelian QHs~\cite{Mazza_PRA.82.043629, Kapit_Mueller_PRL.108.066802, Liu_Kapit_PRB.88.205101, Hafezi_PRB.90.060503, Manna_Nielsen_PRB.98.165147}.

\begin{acknowledgements}
Discussions with L. Mazza, N. Regnault and P. Roushan, and technical support from L. Parisi, are warmly acknowledged. E.M. also thanks M. Ra$\check{\text{c}}$i$\bar{\text{u}}$nas and F. N. {\"U}nal for fruitful discussions and for sharing some of the data presented in Ref.~\cite{Raciunas_Eckardt_PRA.98.063621}.
E.M., T.C. and I.C. acknowledge financial support from the Provincia Autonoma di Trento and from Google via quantum NISQ 
award.
I.C. also acknowledges financial support from the FET-Open Grant MIR-BOSE (737017) and Quantum Flagship Grant PhoQuS (820392) of the European Union.
R.O.U. has been supported by the BAGEP Award of the Science Academy (Turkey).
S.M. acknowledges support from the DFG via TWITTER, the Italian PRIN 2017, and the EU via the PASQUANS and the QuantERA-QTFLAG projects. M.R acknowledges support from the DFG via Grant RI 2345/2-1.
Part of the numerical calculations in this work has been performed with the computational resources provided by the Mogon cluster at Johannes Gutenberg University, Mainz (\href{https://hpc.uni-mainz.de}{hpc.uni-mainz.de}) made available by the CSM and AHRP.
\end{acknowledgements}

\appendix

\section{Benchmark of TTN ansatz}
\label{app:ED}

To perform a systematic benchmark of the TTN results against exact diagonalization (ED), we consider $N=3$ particles on a $8 \times 8$ lattice with periodic boundary conditions (PBCs), and with $\alpha = 1/8$. For these parameters, the magnetic filling is such that the system hosts two QHs. To localize these two QHs at the same position, we also include a five-sites plus-shaped pinning potential centered at site $(4,4)$, with a repulsion $V_i/t=1$ on each site.

\begin{table}[t]
\centering
\begin{tabular}{c c}
State & $E/t$ \\
\hline
$\Psi_0$ & -9.865430303 \\
$\Psi_1$ & -9.860769293 \\
$\Psi_2$ & -9.828784649 \\
\end{tabular}
\caption{Total energy of the three lowest-energy states of $N = 3$ particles on a $8\times 8$ periodic lattice with $\alpha=1/8$, in the presence of a plus-shaped pinning potential -- see text.}
\label{table:HH:energies}
\end{table}

\begin{figure}[b]
\centering
\includegraphics[width=0.95\linewidth]{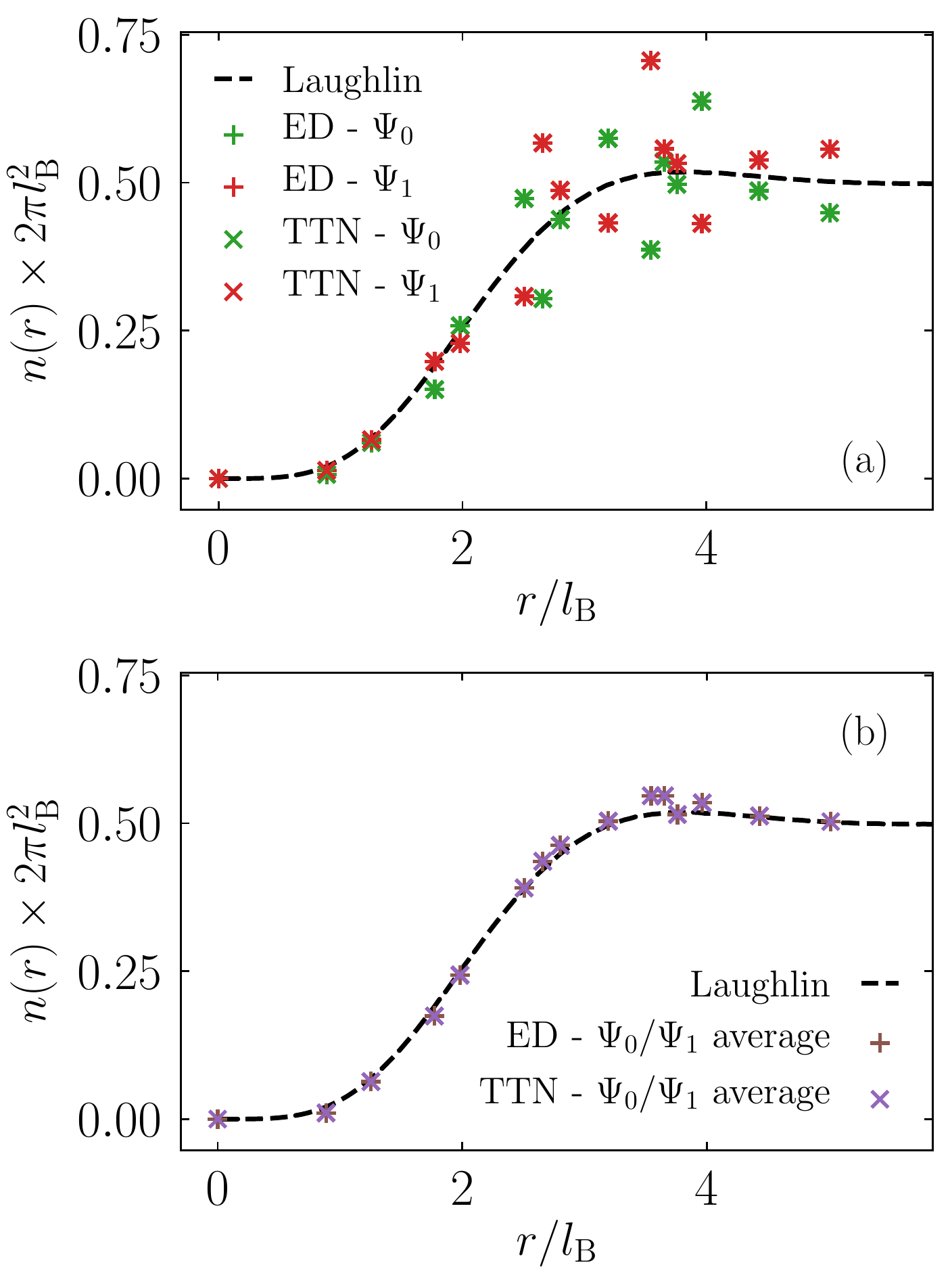}
\caption{(a) Density profiles for the two states $\Psi_0$ and $\Psi_1$ [see text], computed via ED and TTN. (b) Density profiles averaged over $\Psi_0$ and $\Psi_1$, both for TTN and ED. In both panels, the dashed black line is the result for a Laughlin QH state in continuum space.}
\label{fig:ED:depletion}
\end{figure}

Differently from the OBC cases treated in the main text, the toroidal geometry of a PBC system induces a $1/\nu$ topological degeneracy of the ground state~\cite{Wen_PRB.41.9377}, that is, a twofold degeneracy for the case treated in this work. This degeneracy is preserved when the system hosts localized Abelian QHs (up to finite-size effects), in contrast to the general case of non-Abelian or non-localized QHs, in which the number of (quasi-)degenerate states is typically larger~\cite{Regnault_Bernevig_PRX.1.021014, Umucalilar_PRA.98.063629}.

For $N = 3$ and $L = 8$, the Hilbert space dimension is $\approx 4.2 \times 10^4$, and we can obtain the three lowest-energy states $\Psi_0$, $\Psi_1$ and $\Psi_2$ through the ED technique (via the Lanczos algorithm). The corresponding energies [see Table~\ref{table:HH:energies}] display a signature of the (quasi-)degeneracy of $\Psi_0$ and $\Psi_1$, while the state $\Psi_2$ has a larger energy gap -- see Ref.~\cite{Gerster_Montangero_PRB.96.195123} for a study of the spectrum of larger systems.

For this small system, the TTN ansatz has moderate requirements in terms of the bond dimensions of its tensors. In general, within the hierarchical tree structure of a loop-free TTN state, the bonds of a tensor on the $l$-th layer may have dimension up to $d^{2^l}$~\cite{Gerster_Montangero_PRB.90.125154, Silvi_Montangero_SciPost.10.21468}, where $d$ is the dimension of the local Hilbert space. To make this ansatz tractable, one typically introduces a cutoff $D$ on the bond dimension, so that the dimension of each bond becomes $\min(d^{2^l}, D)$. For $L=8$, the root tensor (i.e. the one at the top of the tree structure) has a maximum bond dimension of $\min(d^{2^5}, D) = \min(d^{32}, D)$.

For the hard-core Bose-Hubbard model treated in this work, the local dimension is $d = 2$. By considering a fixed number $N$ of bosons, the number of states is reduced, and we can compute the bond-dimension $D_\mathrm{exact}$ needed to span the entire many-body Hilbert space. For $N = 3$ and $L = 8$, a TTN state with $D \geq D_\mathrm{exact} = 66$ can reproduce an arbitrary many-body state. We then set $D = 66$ in this Appendix, and show that the optimization of the TTN parameters yields the same physical results as ED.

First of all, the TTN energies for $\Psi_0$, $\Psi_1$ and $\Psi_2$ perfectly match with the ED values in Table~\ref{table:HH:energies}, up to the precision reported therein. We also compute the density profile close to the pinned QHs, as shown in Fig.~\ref{fig:ED:depletion} for $\Psi_0$ and $\Psi_1$. These two degenerate states have different profiles, but once we average over the two states we obtain a smooth curve, as known from Ref.~\cite{Liu_Regnault_PRB.91.045126}. Also for the profiles, as for the energies, we observe full agreement between ED and TTN results. This further validates the TTN method, also in the presence of a localized QH excitation.

\section{Bond dimension dependence of observables}
\label{app:TTN_convergence}

In App.~\ref{app:ED}, we showed that for a small system one can reach the saturation of the bond-dimension cutoff ($D = D_\mathrm{exact}$), which makes the TTN ansatz a general parametrization for any state in the Hilbert space. The computational cost of this task, however, becomes prohibitively large for larger systems. For Case I (Case II) in the main text, such saturation would require a bond dimension $D \gtrsim 6\times10^9$ ($D \gtrsim 2\times10^{13}$). Since this is clearly beyond reach, the TTN ansatz can only represent states in the subset of Hilbert space with low to moderate entanglement content, which introduces a bias in the computed observables.

As a representative example, in this Appendix we consider the parameters of Case II reported in Table~\ref{table:HH_parameters}, supplemented with a plus-shaped pinning potential of intensity $V_i/t=1$. In this setup the system has a localized double QH -- see orange diamonds in Fig.~\ref{fig:depl_and_phi_CaseII}(a).
We first look at the energy, for bond dimension $D$ between 100 and 500 -- see data in Table~\ref{table:E_vs_bonddimension}. Even if the large-$D$ convergence is not reached, the energies for the two largest bond dimensions are very close, with a relative energy difference as small as $1.5 \times 10^{-4}$. This suggests that the TTN ansatz with the currently available bond dimension may be sufficiently accurate to represent the ground state of the model under study.
Note that the relative energy differences reached in this work are comparable with the TTN analysis for the case of a homogeneous FCI~\cite{Gerster_Montangero_PRB.96.195123}.

\begin{table}[htb]
\centering
\begin{tabular}{c c}
$D$ (bond dim.) & $E/t$ \\
\hline
100 & -47.508 \\
200 & -47.616 \\
300 & -47.631 \\
400 & -47.654 \\
500 & -47.661 \\
\end{tabular}
\caption{Dependence of the total energy $E$ on the bond dimension $D$, for a double QH in Case II -- see text.}
\label{table:E_vs_bonddimension}
\end{table}

\begin{figure}[b]
\centering
\includegraphics[width=0.9\linewidth]{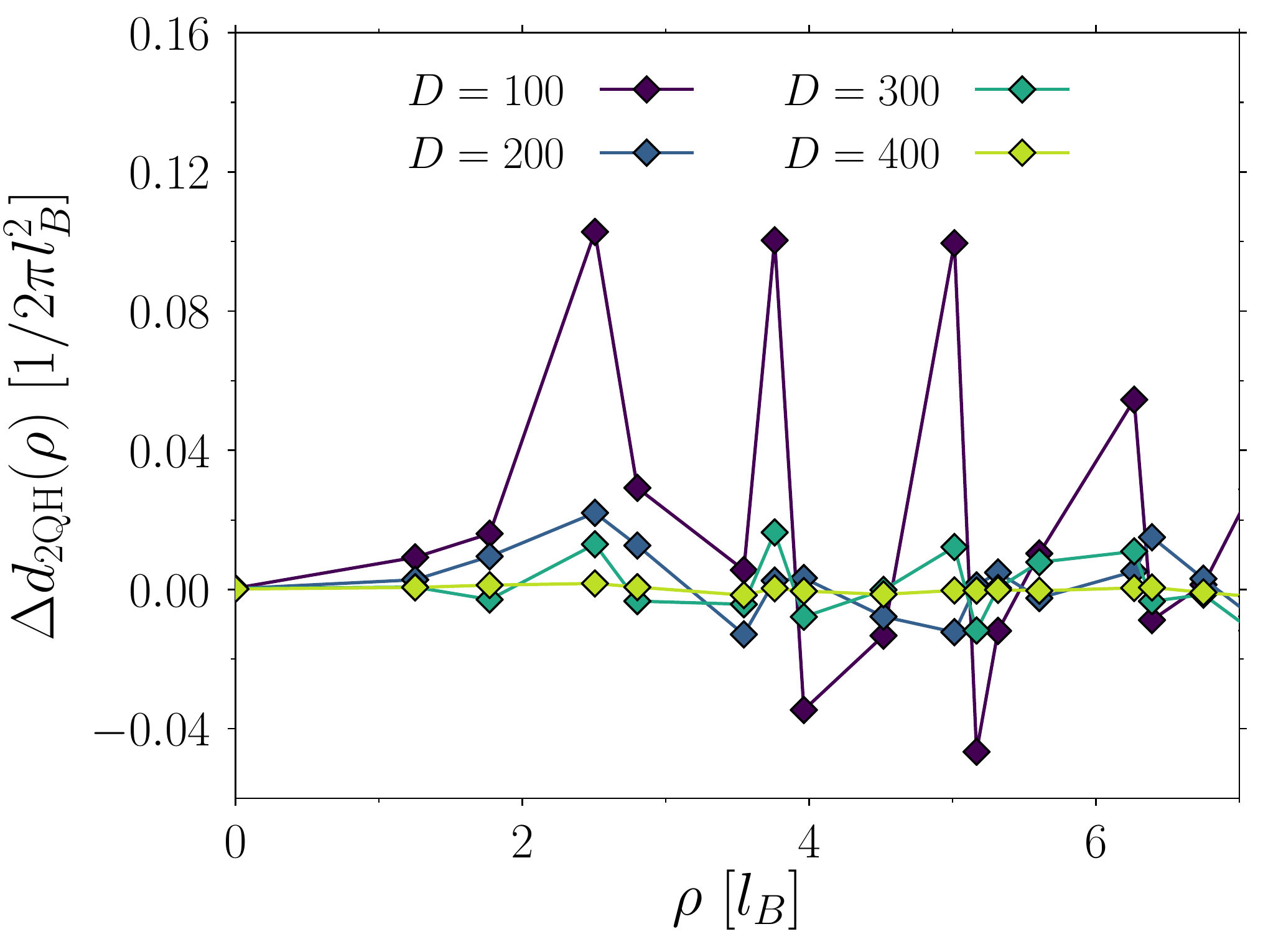}
\caption{Finite-$D$ deviations in the depletion profile of a double
QH in Case II -- see Eq.~\eqref{eq:Delta_d_rho}. The reference value $\Delta d_\mathrm{2QH}(\rho) = 0$ corresponds to $D=500$ [see orange diamonds in Fig.~\ref{fig:depl_and_phi_CaseII} (a)].}
\label{fig:TTN:depletion_convergence}
\end{figure}

Furthermore, we look at how the depletion profile around a QH (the key observable in the current work) depends on the TTN bond dimension. We look again at the same case (a double QH in Case II, with a plus-shaped pinning potential), and compute the depletion profile $d^{(D)}_\mathrm{2QH}(\rho)$ for different values of $D$.
To visualize the dependence on the bond dimension, we take $D=500$ as a reference case and define the depletion-profile deviation as
\begin{equation}
\Delta d_\mathrm{2QH}(\rho) = 
d^{(D)}_\mathrm{2QH}(\rho) - d^{(500)}_\mathrm{2QH}(\rho).
\label{eq:Delta_d_rho}
\end{equation}
This quantity is shown in Fig.~\ref{fig:TTN:depletion_convergence}, for $D$ between $100$ and $400$. We observe that its fluctuations decrease for increasing bond dimension, and that the curve for $D=400$ is barely distinguishable from 0, on this scale. We conclude that for a TTN with $D=500$ (the bond dimension used for Case II in the main text) the systematic error in the depletion profile is negligible.


%

\end{document}